\documentclass[fleqn,twoside]{article}

\usepackage{espcrc2,amssymb}
\usepackage{epsfig,here}
\usepackage[figuresright]{rotating}

\newcommand{\AmS}{{\protect\the\textfont2
  A\kern-.1667em\lower.5ex\hbox{M}\kern-.125emS}}
\title{Effects of neutrino oscillations on the supernova signal in LVD}

\author{M.Aglietta\address[TO]{Institute of Cosmo-Geophysics, CNR, Torino,
University of Torino and INFN-Torino, Italy},
P.Antonioli\address[BO]{University of Bologna and INFN-Bologna, Italy},
G.Bari\addressmark[BO],
C.Castagnoli\addressmark[TO],
W.Fulgione\addressmark[TO],
P.Galeotti\addressmark[TO],
PL.Ghia\addressmark[TO],
P.Giusti\addressmark[BO],
E.Kemp\address[BR]{University of Campinas, Campinas, Brazil},
A.S.Malguin\address[INR]{Institute for Nuclear Research, Russian Academy of
Sciences, Moscow, Russia},
G.Nurzia\address[AQ]{INFN-LNGS, and University of L'Aquila, Italy},
A.Pesci\addressmark[BO],
P.Picchi\addressmark[TO],
I.A.Pless\address[MIT]{Massachusetts Institute of Technology, Cambridge, USA},
V.G.Ryasny\addressmark[INR],
O.G.Ryazhskaya\addressmark[INR],
G.Sartorelli\addressmark[BO], 
M.Selvi\addressmark[BO],
C.Vigorito\addressmark[TO],
F.Vissani\addressmark[AQ],
L.Votano\address[LNF]{INFN-LNF, Frascati, Italy},
V.F.Yakushev\addressmark[INR],
G.T.Zatsepin\addressmark[INR],
A.Zichichi\addressmark[BO]}

\begin{document}

\begin{abstract}
We study the impact of neutrino oscillations
on the supernova neutrino signal 
in the Large Volume Detector (LVD).
The number of expected events for a galactic supernova ($D=10$ kpc)
is calculated, assuming neutrino masses and mixing that  
explain solar and atmospheric neutrino results.
The possibility to detect neutrinos in different channels
makes LVD sensitive to different 
scenarios for $\nu$ properties, such as
normal or inverted $\nu$ mass hierarchy, and/or
adiabatic or non adiabatic MSW resonances associated to $U_{e3}.$ 
Of particular importance are the charged current (c.c.) reactions on 
$^{12}{\rm C}$: oscillations increase by almost one order of 
magnitude the number of events expected from this channel.
\vskip-3.5ex
\end{abstract}

\maketitle

\section{Introduction}
In spite of the lack of a 
``standard'' model of the gravitational
collapse of a massive star, some features of its dynamics and, in
particular, of the correlated neutrino emission
appear to be well established.
At the end of its burning phase a massive star 
($M \gtrsim 8 M_{\odot}$) explodes 
into a supernova (SN), originating a neutron star which cools emitting 
its binding 
energy $E_B\sim 3\cdot10^{53}$ erg mostly in neutrinos.  
The largest part of this energy, almost equipartitioned among 
neutrino and antineutrino species, is emitted in the cooling phase:
$E_{\bar\nu_e} \sim E_{\nu_e} \sim E_{\nu_x} \sim E_B/6$
($\nu_x$ denotes generically $\nu_\mu,\bar{\nu}_\mu,\nu_\tau,\bar{\nu}_\tau$).
The energy spectra are approximatively Fermi-Dirac, but 
with different temperatures, 
since $\nu_e,$ $\bar{\nu}_e$ and $\nu_x$ have 
different couplings with the stellar matter:
$T_{\nu_e}<T_{\bar\nu_e}<T_{\nu_x}.$ 
These features are common to all existing stellar collapse 
models (see for instance \cite{BUR92,JAN96,MEZ01,RAF01,TOTAN}), 
and lead to rather model independent expectations for 
supernova neutrinos. The observable signal is
then sensitive to intrinsic $\nu$ properties, as oscillation of massive 
neutrinos. Indeed, as we demonstrate below, 
oscillations change significantly the expected 
number of events in LVD. 

The paper is organized as follows: 
in Sect.2 we briefly describe the LVD detector, with particular attention
to the detectable neutrino reactions.
In Sect.3 we recall the main features of MSW conversion \cite{MSW,WOL}
in the stellar matter, focusing on the LMA solution,
favored by the most recent results on solar 
neutrinos. For a supernova occurring at $D=10$ kpc,
with pure Fermi-Dirac $\nu$ energy spectra,
we calculate the number of events expected
in LVD, in the different interactions channels.
Sect.4 is devoted to conclusions and discussion of the results.

\section{The LVD experiment and SN neutrino interactions}
The Large Volume Detector (LVD) in the INFN Gran Sasso National Laboratory,
Italy, consists of an array of 840  
scintillator counters, 1.5 m$^3$ each. These are 
interleaved by streamer tubes, and arranged in a compact and modular 
geometry; a detailed description is in \cite{LVD}.
The active scintillator mass is $M=1000$ t. 
There are two subsets of counters: the
external ones ($43 \%$),
operated at energy threshold ${\cal E}_h\simeq 7$ MeV,
and inner ones ($57 \%$), better shielded from rock radioactivity and 
operated at ${\cal E}_h\simeq 4$ MeV. 
In order to tag the delayed $\gamma$ pulse due to
$n$-capture, all counters are equipped with an additional discrimination
channel, set at a lower
threshold, ${\cal E}_l\simeq 1$ MeV.
Relevant features of the detector are: $(i)$ good event localization; $(ii)$
accurate absolute and relative timing:
$\Delta\mathrm{t}_{\rm abs}=1\, \mu \rm{s}$, 
$\Delta\mathrm{t}_{\rm rel}=12.5\, \rm{ns}$;
$(iii)$ energy resolution: 
$\sigma_{E}/{E} = 0.07 + 0.23\cdot ({E} /\rm{MeV})^{-0.5}$.

The observable neutrino reactions are: 
\\ \noindent {\em (1)}\  $\bar\nu_e p,e^+ n$, observed
through
a prompt signal from ${e}^+$ above threshold ${\cal E}_h$
(detectable energy $E_d 
\simeq E_{\bar\nu_{e}}-1.8$ MeV $+ 2 m_e c^2 $),
followed by the signal
from
the ${n p,d} \gamma$ capture ($E_{\gamma} = 2.2$ MeV),
above ${\cal E}_l$ and 
with a mean delay  $\Delta t \simeq 180~\mu \mathrm{s}$. 
\\ \noindent {\em (2)}\  
$\nu_e\, ^{12}\mathrm{C},^{12}\!\mathrm{N}\ e^-$, 
observed through two signals: 
the prompt one due to the $e^-$ above ${\cal E}_h$
(detectable energy $E_d \simeq E_{\nu_e}-17.8$ MeV)
followed by the signal, above ${\cal E}_h$, from the $\beta^+$ decay of 
$^{12}\!\mathrm{N}$ (mean life time $\tau = 15.9$ ms).
\\ \noindent {\em (3)}\  
$\bar\nu_e\, ^{12}\rm{C},^{12}\!\rm{B}\ e^+$, 
observed through two signals:
the prompt one due to the $e^+$
(detectable energy $E_d \simeq E_{\nu_e}-13.9\, \mathrm{MeV} 
+ 2m_e c^2$)
followed by the signal
from the $\beta^-$ decay of $^{12}\!{\rm B}$
(mean life time $\tau= 29.4$ ms). As for reaction {\em (2)}, the
second signal is detected above the threshold ${\cal E}_h$.
\\ \noindent {\em (4)}\  
$\stackrel{\scriptscriptstyle (-)}{\nu}_{\!\ell}\, 
{}^{12}{\rm C},\stackrel{\scriptscriptstyle (-)}{\nu}_{\!\ell}\, 
{}^{12}{\rm C}^*$ 
($\ell=e,\mu,\tau$), whose signature  
is the monochromatic photon 
from carbon de-excitation ($E_{\gamma}=15.1$ MeV),
above ${\cal E}_h.$
\\ \noindent {\em (5)}\  
$\stackrel{\scriptscriptstyle (-)}{\nu}_{\!\ell}\, 
e^-,\stackrel{\scriptscriptstyle (-)}{\nu}_{\!\ell}\, 
e^-$, which yields
a single signal, above ${\cal E}_h$, due to the recoil electron.

\section{Neutrino mixing: events in LVD}
In the study of supernova neutrinos,
$\nu_{\mu}$ and $\nu_{\tau}$ are indistinguishable, both in the star and in
the detector; consequently, in the frame of three-flavor oscillations,
the relevant parameters are just
$(\Delta m^2_{{\rm sol}}, U_{ e2}^2)$ and 
$(\Delta m^{2}_{\rm atm}, U_{e3}^2)$. 
We will adopt the following numerical values:
$\Delta m^2_{{\rm sol}}=5 \cdot 10^{-5}{\rm eV}^2$, 
$\Delta m^{2}_{\rm atm}=2.5 \cdot 10^{-3} {\rm eV}^2$, 
$U_{e2}^2=0.33;$ the selected solar parameters 
$(\Delta m^2_{{\rm sol}}, U_{e2}^2)$ describe a
LMA solution, favored by recent analyses \cite{Fogli}.\\
For a normal mass hierarchy scheme,  
neutrinos (not anti-neutrinos) cross two resonance layers:
one at higher density (H), which corresponds to $\Delta m^{2}_{\rm atm}$, and
the other at lower density (L), corresponding to 
$\Delta m^{2}_{{\rm sol}}$.\footnote{
For inverted mass hierarchy, transitions at
the higher density layer occur in the anti-neutrino 
sector, while at the lower density layer they occur 
in the neutrino sector. Anyway, 
both in case of normal and inverted mass hierarchy,
the dynamics of collapse is not affected, 
since these layers are located far outside the core of 
the star.}
Given the energy range of supernova neutrinos
($5\ {\rm MeV}\lesssim E_{\nu} \lesssim 50$ MeV),
and considering a star density profile $\rho \propto 1/r^3$, 
the adiabaticity condition is always satisfied at the L resonance
for any LMA solution, while at the H resonance, this 
depends on the value of $U_{e3}^2$. 
When $U_{e3}^2 \gtrsim 5 \cdot 10^{-4}$ 
the conversion is completely adiabatic, meaning that 
$\nu_e$ are completely converted into the mass
eigenstate $\nu_3$ (detected at the Earth mainly as
$\nu_{\mu}$ and $\nu_{\tau}$).
Therefore, the SN neutrino signal could feel the effect of $U_{e3}^2$
(and could also help to discriminate the type of mass hierarchy)
\cite{Panta}, \cite{Dutta}, \cite{Dighe}.\\
We calculated the number of events
expected in the reaction channels {\em (1)-(4)} in
LVD,\footnote{
The signature of channel {\em (5)}
is not as clear as the other ones,
and the number of expected events is low; 
therefore, we disregard it in the following.
}
in the cases of no-oscillation and oscillation, 
under the following hypotheses:
\\ \noindent $\star$ We assumed
a supernova exploding at $D=10$ kpc, with an energy release
$E_{\rm tot}= 3\cdot 10^{53}$ erg, pure Fermi-Dirac time integrated spectrum,
energy equipartition,  
and neutrinospheres temperatures as
$T_{\nu_e} = T_{\bar\nu_e} = T_{\nu_{x}}/2$.
\\ \noindent $\star$ We included the active mass of the detector 
and the energy thresholds, as described in Sect.2.
We used the following values of detection efficiencies 
above threshold:
$\epsilon_{\bar\nu_e\, p,e^+\, n} = 95\% $ and
$\epsilon_{n\, p,d\, \gamma_{2.2}}=50\% $;
$\epsilon_{\nu_e\, {\rm C},{\rm N}\, e^-}= 85\% $;
$\epsilon_{\bar\nu_e\, {\rm C},{\rm B}\, e^+}= 70\% $;
$\epsilon_{\nu_\ell\, \rm{C},\nu_\ell\, \rm{C}}= 55\% $ 
\cite{ANT91}.
\\ \noindent $\star$ In the oscillation case, 
we used two extreme values for $U_{e3}^2$:
$U_{e3}^2=10^{-2}$ and $U_{e3}^2=10^{-6}$,
and the above mentioned mixing
parameters (normal mass hierarchy, LMA solution).
\\ \noindent $\star$ We did not include Earth matter effects 
(``open sky'' neutrino burst).\\
\begin{figure}[htb]
\vspace{-0.5cm}
\begin{center}
\vskip-5mm
\mbox{\epsfxsize=7cm \epsfbox{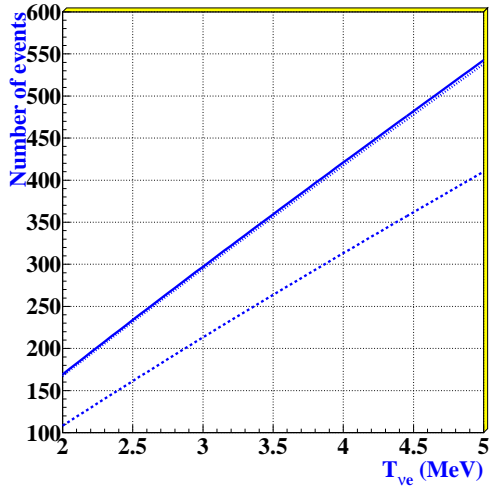}}
\vskip-10mm
\caption{Number of events expected in LVD, in the reaction
$\bar\nu_e p,n e^+$, as a function of 
$T_{\bar\nu_e}\equiv T_{\nu_e} $:
the dashed line represents the no-oscillation case, while full and dotted
lines represent the oscillation case, adiabatic and non adiabatic,
respectively.}
\label{fig:ibd}
\vskip-10mm
\end{center}
\end{figure}
Fig.1 shows the number of expected events versus $T_{\nu_e}$ 
in the inverse $\beta$ decay $\bar\nu_e$ reaction:
a large increase due to $\nu$ mixing is clearly visible, with respect to
the no-oscillation case.
It should be noted that the number of $\bar\nu_e p$ 
events is practically the same both for adiabatic and
non-adiabatic conditions, since, for normal
mass hierarchy, MSW effect takes place 
in the neutrino sector only. 
Quite a different picture would appear, if we were to 
assume inverse mass hierarchy.

\begin{figure}[htb]
\begin{center}
\vskip-3mm
\mbox{\epsfxsize=7cm \epsfbox{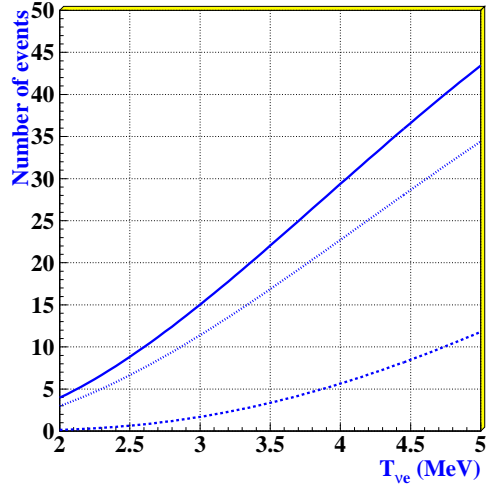}}
\vskip-10mm
\caption{Number of events expected in LVD, in c.c.\ 
interactions with $^{12}{\rm C}$ as a function of 
$T_{\bar\nu_e}\equiv T_{\nu_e} $:
the dashed line represents the no-oscillation case, while full and dotted
lines represent the oscillation case, adiabatic and non adiabatic,
respectively.}
\label{fig:ccc}
\vskip-10mm
\end{center}
\end{figure}

Fig.2 shows the expected 
total number of c.c. interactions with $^{12}{\rm C}$, due to both 
$\nu_e$ and $\bar\nu_e.$\footnote{
Since mean life times of $\beta^\pm$ decay are similar (see
Sect.2), $\nu_e$ and $\bar\nu_e$ are distinguishable 
only on statistical basis.
Note that, at $T=4$ MeV, 
we expect 6 events due to 
$\bar\nu_e\, ^{12}{\rm C},^{12}\!{\rm B}\ {e}^+$ in both cases with
oscillations.
}
The mixing results in an increase of the number of events,
either for adiabatic or for non adiabatic conditions: 
in case of adiabaticity 
the increase is larger, and this 
is solely due to $\nu_e$ interactions.
\begin{figure}
\begin{center}
\vskip-3mm
\mbox{\epsfxsize=7cm \epsfbox{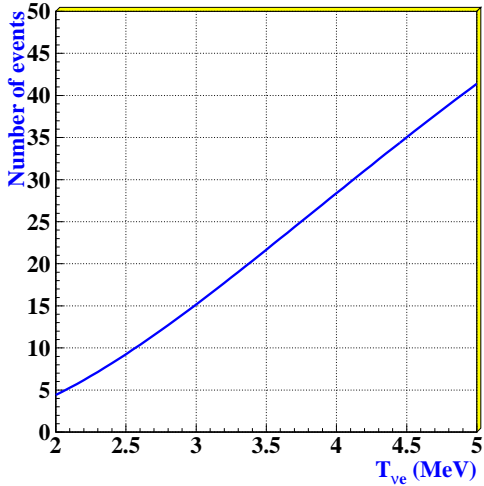}}
\vskip-10mm
\caption{Number of events expected in LVD, 
in n.c.\ interactions with $^{12}{\rm C}$
as a function of $T_{\nu_e}$.}
\label{fig:ncC}
\vskip-10mm
\end{center}
\end{figure}

Finally, the expected number of events in neutral currents (n.c.) interactions
with $^{12}{\rm C}$ is shown in Fig.3:
they are of course
insensitive to $\nu$ mixing. However, the
number of carbon de-excitations can test the temperature
of neutrinospheres at the source \cite{Fulg}, and therefore could be
used in combination with c.c.\ data to overcome theoretical
uncertainties on the temperature.

\section{Conclusions and discussion}

The observation of a neutrino burst due to the
explosion of a galactic supernova can add precious information about
neutrino mass and mixing scenarios, in a complementary way with respect to
solar, atmospheric and terrestrial $\nu$ experiments.

We have studied the signal at LVD from a  
SN exploding at $D=10$ kpc for
3-flavor $\nu$ oscillation, assuming 
the LMA-MSW solution for solar $\nu$ and 
normal mass hierarchy. We calculated 
the expected number of events 
for extreme values of $U_{e3}^2.$
Varying oscillation parameters, 
we found an increase up to 50\% of the signal due to inverse
$\beta$ decay,  
and an increase by almost one order of magnitude of the 
signal due to c.c.\  reactions on carbon. 
We remind the reader that the signatures
of these reactions  in LVD are very clear.

We plan to extend the calculation to include
more refined energy and time distributions of $\nu$ 
and the Earth matter effect.
The latter leads to peculiar distortions of 
the spectra of SN $\bar{\nu}_{{e}}$ and $\nu_{{e}},$ 
depending on their trajectory in the Earth,\footnote{Even 
if we should not detect the optical counterpart of 
the next galactic supernova, the $\nu$
signal itself can inform us on the source position 
by the directional (anti)neutrino electron 
scattering as observed by \v{C}erenkov detectors.}
on mass hierarchy and on adiabaticity of H resonance. 
Earth matter effect amplifies the difference between the 
two cases with oscillation in Fig.2, due to 
$\nu_e$ regeneration in the non-adiabatic case.

The combination of well separated classes of events, namely 
those due to charged and neutral current interactions on carbon
and those due to inverse $\beta$ decay,  
can help to distinguish between different 
scenarios of massive neutrinos and astrophysical 
parameters.
In addition, as recently discussed \cite{Lunardini2001}, 
the comparison of the results of a {\em network} of detectors could 
permit us to fully exploit the supernova neutrino signal, thus learning
on neutrino intrinsic properties. Indeed, we could finally confirm the 
solution of the solar neutrino problem, learn on the mixing ${U}_{e3},$ 
and identify the hierarchy of the neutrino mass spectrum.   

\section{Acknowledgements}                   
This work was initiated by a conversation with Sandro Bettini, to whom we
are grateful.\\
We would like to acknowledge the Gran Sasso Laboratory Staff for the 
continuous and valuable support.

\end{document}